\documentclass[preprint]{aastex}
\usepackage{amsmath}
\usepackage{hyperref}

\usepackage[english]{babel}
\usepackage{graphicx}
\usepackage{latexsym}
\usepackage{amsmath,amssymb}
\usepackage{natbib}
\usepackage{dcolumn}
\usepackage{bm}
\usepackage{calrsfs}
\usepackage{ulem}
\usepackage[usenames]{color}
\usepackage{xspace}

\newcommand{\nn}{\mbox{} \nonumber \\ \mbox{} }
\newcommand{\ba}{\begin{eqnarray}}
\newcommand{\ea}{\end{eqnarray}}

\newcommand{\Alfven}{ Alfv\'{e}n }

\newcommand{\Bf}{{magnetic field}}

\newcommand\eg{\textit{e.g.}}
\newcommand\cf{\textit{cf.\ }}

\def\be{\begin{equation}}
\def\ee{\end{equation}}

\begin{document}

\title{Cosmic ray-modified shocks:  appearance of an isothermal jump}
\author{Maxim Lyutikov}
\affil{ Department of Physics  and Astronomy, Purdue University, 
 525 Northwestern Avenue,
West Lafayette, IN
47907-2036, USA; lyutikov@purdue.edu
}

%%%%%%%%%%%%%%%%%%%%%%%%%%%%%%%%%%%%%%%%%%%%%%%%%%
\begin{abstract}
We point out that 
for sufficiently strong shocks, with Mach number $ M_1 >  \sqrt{\frac{3 \gamma -1}{(3-\gamma ) \gamma }}=  1.34$ ($\gamma=5/3$), the solutions for cosmic ray-modified shocks experiences a bifurcation. As a result, for super-critical flows  an isothermal jump forms (which is  {\it not} a shock). The isothermal jump forms due to  the energy diffusion of  fast, but energetically subdominant cosmic rays.  For super-critical flows the isothermal jump  appears regardless of a particular feed-back mechanism from the CRs.  The compression ratio at the  isothermal jump is $2/ (\gamma-1)=3$, so that in the test particle regime the expected spectrum of low energy CRs  experiencing   first-order Fermi process is 
$p = 2\gamma/(3- \gamma) = 5/2$, steeper than conventional $p=2$.
\end{abstract}

\section{Cosmic rays' feedback on shock structure}
Cosmic rays  (CRs) modify internal structure of astrophysical shocks \citep[\eg][]{1980ApJ...238..410B,1982A&A...111..317A,1982MNRAS.198..833D,1987PhR...154....1B,2001RPPh...64..429M,2006MNRAS.371.1251A}. \cite{1980ApJ...238..410B} calculated perturbative effects of the CR on the shock structure expanding in powers of small CR pressure \citep[see also][]{1983RPPh...46..973D}.

%Attempts at non-linear feedback \citep[\eg][]{2006MNRAS.371.1251A} hard to start with a particular flow structure. 
%The effects of  CRs on shock structure have nearly universally been treated as a perturbation, starting with the work of  \cite{1980ApJ...238..410B}, who calculated effects of the CR on the shock structure expanding in powers of small CR pressure. \citep[][discussed non-perturbative approach]{1997ApJ...491..584M,2001RPPh...64..429M}. 

The simplest way to calculate the CR ray feedback is within two-fluid model, whereby CR form a separate light, highly diffusive fluid. Here, experice with radiative shocks comes handy.
 It is well known in the theory of radiative shocks \citep[\eg][parag. 95]{LLVI}, see also  \cite{ZeldovichRaizer}, that for sufficiently strong shocks the internal structure of the solution changes qualitatively - in some limits regardless of the strength of the feed-back an isothermal jump forms within the flow. Similar  effect should occur in  CR-modified shocks:  an extended precursor is followed by an isothermal jump, not a sub-shock, as we argue below. Mathematically, addition of CR diffusion leads to a higher order differential equation for the velocity and, thus, cannot be treated as a perturbation.

\section{Non-perturbative CR feedback}
\subsection{The iso-thermal jump}
For sufficiently strong shocks the CR feedback is non-perturbative, as we discuss next. 
The first most important effect on the shock structure from cosmic rays  \citep[similar to effects of radiation in atmospheric explosions,][]{ZeldovichRaizer}  is the diffusive spreading of energy of CRs. This can be seen from the following argument. Strong (initial pressure equals zero) CR-modified shocks in the hydrodynamic approximation obey the following equations
  \ba &&
  \beta_1 \rho_1 = \beta  \rho
  \nn &&
   \rho_1 \beta_1^2 = p_{ tot} + \rho_{ tot} \beta^2
   \nn &&
    \rho_1 \beta_1^3/2=( w_{ tot} +  \rho_{ tot} \beta^2/2) \beta +F_{CR}
     \nn &&
      p_{ tot}= \frac{\rho}{m_p} T + \frac{u_{CR}}{3}
      \nn &&
      w_{ tot}= \frac{\gamma}{\gamma-1} \frac{\rho}{m_p} T+ \frac{4}{3} u_{CR}
    \label{conserv}
    \ea
    where $ p_{ tot}$ and $ w_{ tot}$ are total pressure and enthalpy, composed of plasma and CR contribution, and 
$F_{CR}$ is the energy flux   carried by CRs. In the diffusive approximation $F_{CR} \propto \partial_z  u_{CR}$.   Values on the left refer to the far upstream.  
Thus, cosmic rays contribute to pressure and energy flux. Importantly, CR contribution to pressure is an addition - and thus is small for $u_{CR} \ll p_{gas}$. On the other 
    hand, the term with energy flux $F_{CR}$ changes the order of the differential equation, and hence the structure of the solutions. This is the most important effect.

Thus, the first effects of  CRs on the shock  is the  redistribution energy due to CR diffusion, leaving only  $F_{CR}$ term in (\ref{conserv}). Then at each point  a 1D stationary non-relativistic flow is described by the following set of equations (mass, momentum and energy flux conservation)
 \ba &&
\rho _1 v_1=\rho  v
\nn &&
\rho _1 v_1^2=p+\rho  v^2
\nn &&
\frac{1}{2} \rho _1 v_1^3=F_{CR}+v
   \left(\frac{\rho  v^2}{2}+w\right)
\ea
 \citep[\cf][Sec. VII.3 and Eqns. (7.10), (7.40)]{ZeldovichRaizer}.

   Both far upstream and far downstream the  CR flux is zero. 
Introducing (the inverse of the) compression ratio $\eta = \rho_1/\rho$, the shock jump conditions give
\ba && 
\eta_2 = \frac{\gamma-1}{\gamma+1}
\nn &&
T_2= 2 \frac{\gamma-1}{(\gamma+1)^2} m_p v_1^2
\label{2}
\ea
where subscript $2$ denotes values far downstream.

\begin{figure}[h!]
 \vskip -.1 truein
 \centering
\includegraphics[width=0.9\textwidth]{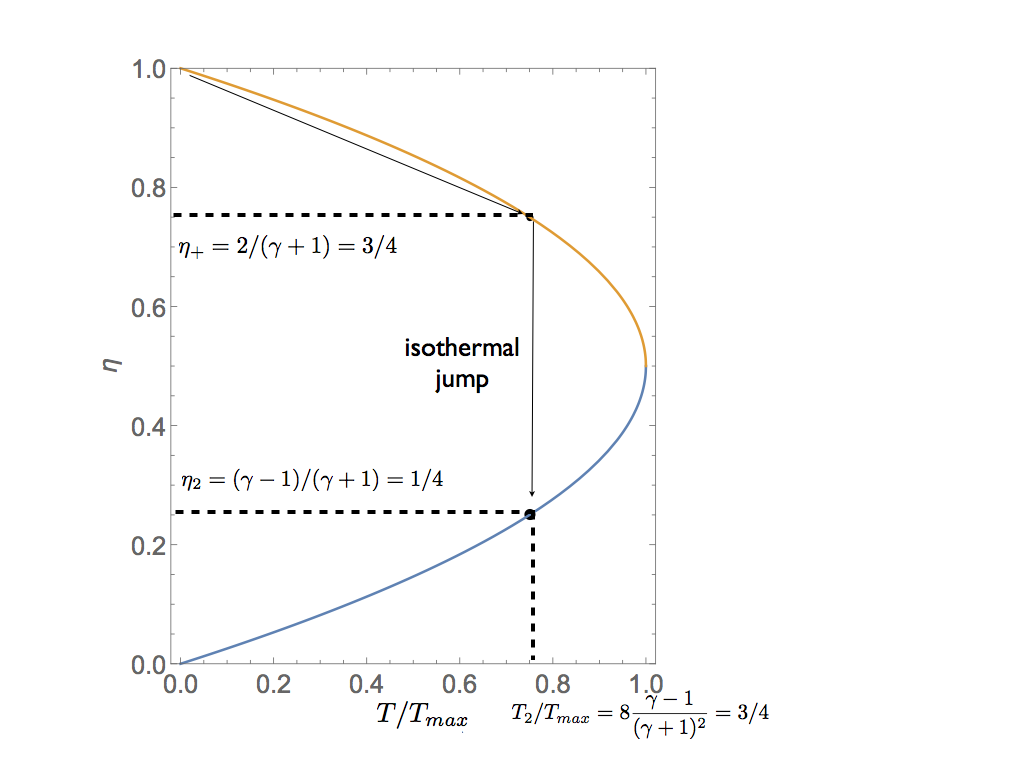}
\caption{ Compression ratio   as function of  temperature for very strong shocks. Two highlighted  points correspond to the jump solutions; only the lower point is physical. To reach the physical solution it is necessary to pass through two special points: unphysical solution corresponding to the same final temperature and a special point at 
$\theta _{T,max}$
}
% \vskip -.5 truein
\label{ofbeta}
\end{figure}

Within the shock, independently of the energy flux equation, the momentum conservation can be written as
%\ba &&
\ba &&
T= (1-\eta) \eta m_p v_1^2
\label{T}
\\ &&
\eta(T) = \frac{1}{2} \pm \sqrt{\frac{1}{4}- \frac{T}{m_p v_1^2}}=  \frac{1}{2}\left(1 \pm \sqrt{1- \frac{T}{T_{max}}}\right).
\label{momentum}
\ea
where  $T_{max} =   m_p v_1^2/4$.  Thus, there are two branches of $\eta(T)$, see Fig. \ref{ofbeta}. It is the upper branch that connects to the pre-shock state with $\eta=1$.
Importantly,  for super-critical shocks the final state (\ref{2}) is located at the lower branch.

Note, that
\be
\frac{T_2}{T_{max}} = 8 \frac{\gamma-1}{(\gamma+1)^2}= \frac{3}{4} < 1
\ee
Thus, as the state evolves along the upper branch, the terminal temperature is reached before the the terminal compression. It is required that temperature increase monotonically \citep[\eg][ Eq. (95.3)]{LLVI}. Thus, since  ${T_2}<{T_{max}}$, the final state cannot be reached continuously. There should be  an isothermal jump at $T=T_2$, Fig. (\ref{ofbeta}).

Note that we have derived the behavior of the compression ratio  (and thus velocity of the flow) as a function of temperature  {\it without specifying a particular CR feedback mechanism!}  How the system evolves toward the iso-thermal jump depends on the particular form of $F_{CR}$, but the existence of the  iso-thermal jump is a consequence of the momentum conservation and total jump conditions, which are independent of $F_{CR}$.

Qualitatively, shock jump conditions without diffusive effects may be written as continuos algebraic relations. Addition of diffusive terms modifies the structure of these relations - instead of algebraic, the energy evolution becomes a differential equation. There  are spacial points in the equation - \eg, stationary solutions   correspond to the shock jump conditions. 
 A continuous solution cannot pass through some of the special points, \eg\ $T= T_{max}$  - this determines the formation of the iso-thermal jump.

\subsection{Appearance of the iso-thermal jump}

The above derivation assumed that the upstream medium is cold, so that the shock is infinitely strong. If the upstream plasma has temperature $T_1$ (so that Mach number is $M_1 = \sqrt{ \frac{m_p}{\gamma T_1}} v_1$), the compression ratio is 
\be
\eta_\pm =\frac{1}{2} \left(\frac{T_1}{v_1^2 m_p}-\sqrt{\left(\frac{T_1}{v_1^2
   m_p}+1\right){}^2-\frac{4 T}{v_1^2 m_p}}+1\right)
  \label{etapm}
\ee
Thus, the maximal temperature is 
\be
T_{max}= \frac{\left(v_1^2 m_p+T_1\right){}^2}{4 v_1^2 m_p}=\frac{v_1^2 m_p \left(\gamma  M_1^2+1\right){}^2}{4 \gamma ^2 M_1^4}
\ee
Post-shock temperature  and compression ratios are
\ba &&
T_2 = -\frac{2 (\gamma -1) \gamma  T_1^2}{(\gamma +1)^2 v_1^2 m_p}+\frac{2 (\gamma -1) v_1^2
   m_p}{(\gamma +1)^2}-\frac{\left(\gamma ^2-6 \gamma +1\right) T_1}{(\gamma +1)^2}
   \nn &&
   \eta_2=\frac{(\gamma -1) v_1^2 m_p+2 \gamma  T_1}{(\gamma +1) v_1^2 m_p}
   \ea
Equating $T_{max}$ to $T_2$ we find that isothermal jump forms for 
\be
T_1 < \frac{(3-\gamma ) v_1^2 m_p}{3 \gamma -1}, \, 
M_1 > M_{crit} =  \sqrt{\frac{3 \gamma -1}{(3-\gamma ) \gamma }}= \frac{3}{\sqrt{5}}= 1.34
\ee
At this point $p_2/p_1= (\gamma+1)/(3-\gamma)$, \cf, \cite{LLVI} Eq. (95.7).

\begin{figure}[h!]
 \vskip -.1 truein
 \centering
\includegraphics[width=0.99\textwidth]{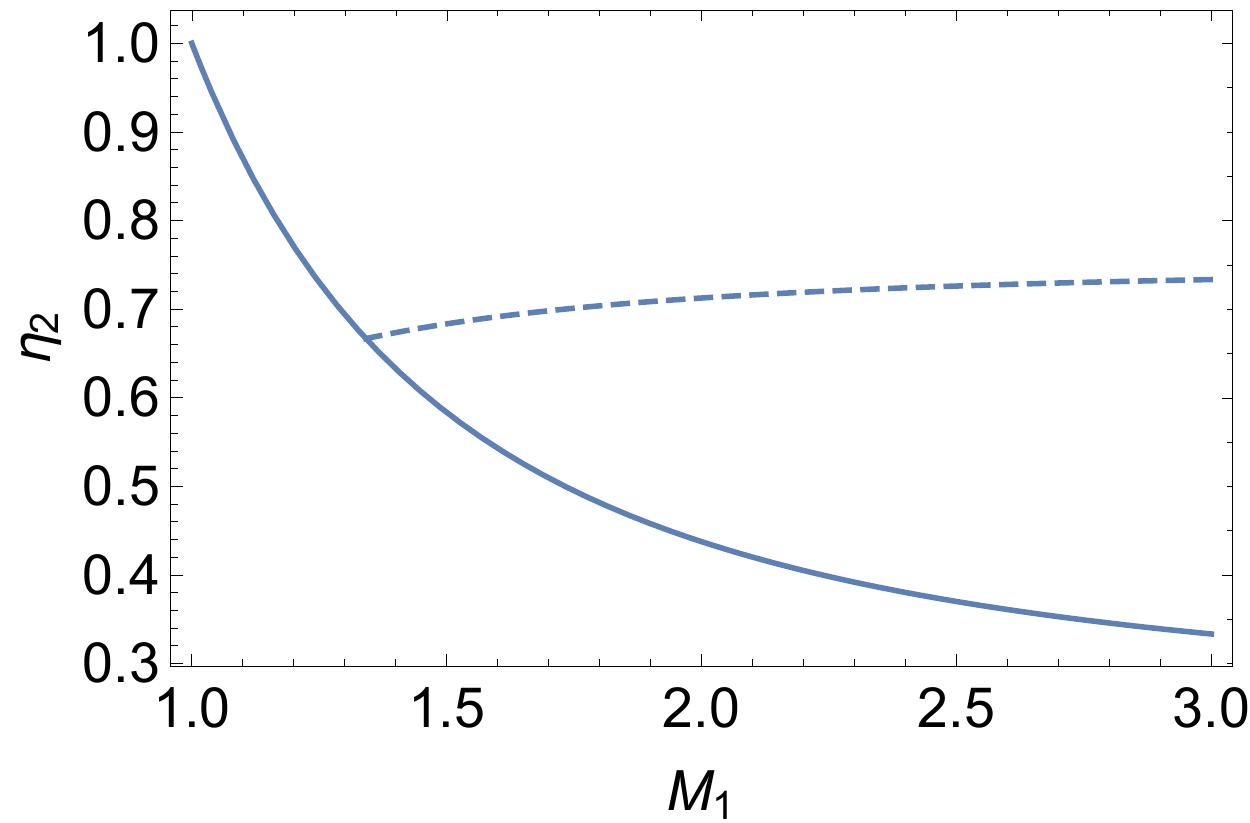}
\caption{ Transition to supercritical shocks as a bifurcation problem.  The final compression ratio $\eta_2$  (solid line) as a function of upstream  Mach number $M_1$.
For $M_1 < M_{crit}=3/\sqrt{5}$ the final solution is reached continuously, while for $M_1 > M_{crit}$  there is bifurcation of solutions (dashed line), so that the final state is reached though an isothermal jump. }
% \vskip -.5 truein
\label{bifurcation003}
\end{figure}

We also  point out that the transition through critical Mach number can be viewed as a bifurcation problem. For $M _1 < M_{crit}$ we have one branch,
\be
\eta = \frac{(\gamma -1) M_1^2+2}{(\gamma +1) M_1^2}
\ee
while for  $M _1 >  M_{crit}$ there is another branch
\be
\eta = 
\frac{\gamma  \left(2 M_1^2-1\right)+1}{\gamma  (\gamma +1) M_1^2}, 
\ee
see Fig. \ref{bifurcation003}.

\begin{figure}[h!]
 \vskip -.1 truein
 \centering
\includegraphics[width=0.99\textwidth]{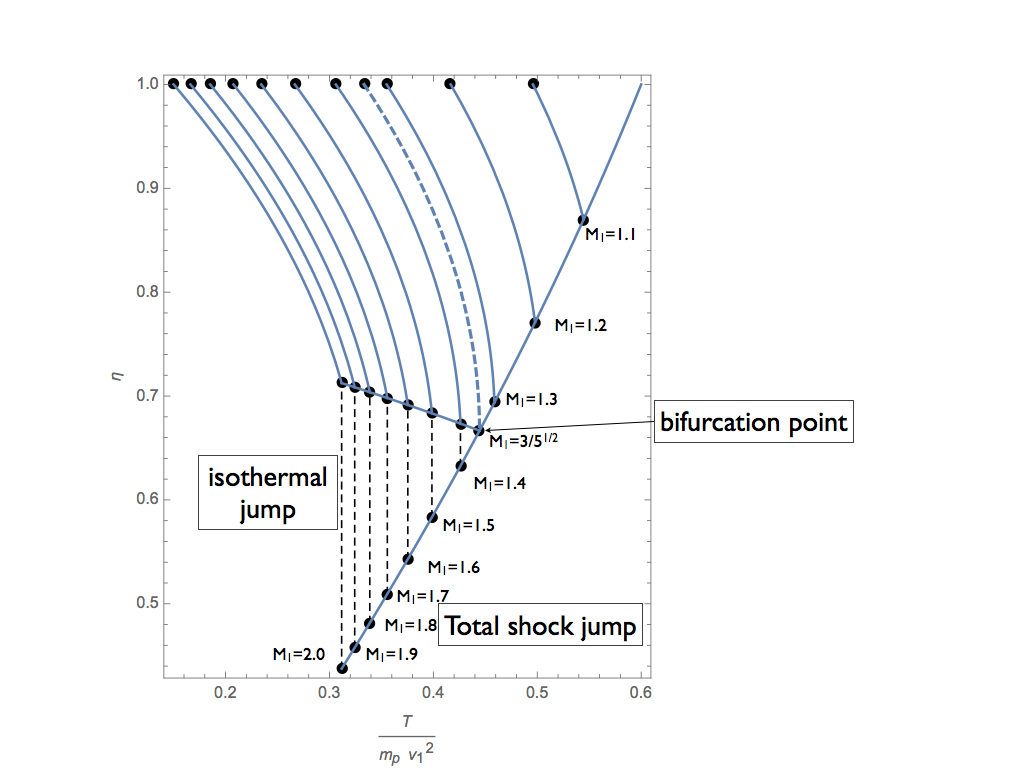}
\caption{Evolution of compression ratio $\eta$ versus temperature (normalized by $m_p v_1^2$ for different Mach  numbers $M=1.1, \,  1.2...2$). The flow starts at $\eta=1$. For $M_1< M_{crit}=3/\sqrt{5}$ the final solution is reaches in a continuous way. There is a bifurcation point at   $M= M_{crit}$: for larger $M_1$ the final state is reached through an isothermal jump.}
% \vskip -.5 truein
\label{jshock-jump005}
\end{figure}

Evolution of quantities in the flow are depicted in Fig. \ref{jshock-jump005}. For $M< M_{crit}=3/\sqrt{5}$ the final solution is reaches in a continuous way. There is a bifurcation point $\{M_1=3/\sqrt{5}, \, \eta=2/3 \}$. For larger $M_1$ the final state is reached through an isothermal jump.

Note that the ratio of the final temperature $T_2$ to maximal temperature $T_{max}$ never exceeds unity:
\ba &&
\frac{T_2}{T_{max}}= 
\frac{-8 (\gamma -3)^2 (\gamma -1) \gamma +4 \left(3 \gamma ^4-28 \gamma ^3+66 \gamma ^2-28
   \gamma +3\right) m_1^2+8 (1-3 \gamma )^2 (\gamma -1) m_1^4}{(\gamma +1)^2 \left(-\gamma
   +(3 \gamma -1) m_1^2+3\right){}^2} \leq 1
   \nn &&
   m_1= \frac{M_1}{M_{crit}}
   \ea
   This ratio reaches unity only at  $m_1=1$. In this case 
   \be
   \eta _{crit}= \frac{1+ \gamma}{3 \gamma-1} = 2/3
   \ee

\subsection{Structure of the precursor}

The appearance of the isothermal jump is independent of the particular form of $F$, but the evolution towards the iso-thermal jump depends on it. To resolve the structure of the precursor one needs to relate  the CR pressure to the fluid parameters.
The energy conservation gives 
\be
\frac{F_{CR}}{n_1 m_p v_1^2} = \frac{(1-\eta ) (-\gamma  (1-\eta )+\eta +1)}{2 (1-\gamma )} =- \frac{(1-\eta ) \left(\eta -\eta _2\right)}{2 \eta _2} \rightarrow
\frac{1}{2} (1-\eta)(1-4 \eta)
\label{energy}
\ee
 For finite $F_{CR}$ (\ref{energy}) and  (\ref{momentum}) -  with a proper form of $F(\eta,T)$ -  determine the structure of the shock.

 As a qualitative example, let us assume that density of CRs follows plasma density, so that
\be
F_{CR}= - \kappa \partial_z \rho
\ee
where $\kappa$ absorbs the diffusion coefficient of CR and the scaling with plasma density. We find
\be
\frac{\kappa}{ v_1^3} \partial_z \eta = -  \frac{(1-\eta ) \eta^2 \left(\eta -\eta _2\right)}{2 \eta _2} \rightarrow
\frac{1}{2} (1-\eta)(1-4 \eta)\eta^2
\label{main}
\ee
where $\eta_2 = (\gamma-1)/(\gamma+1)$.

\begin{figure}[h!]
 \vskip -.1 truein
 \centering
\includegraphics[width=0.99\textwidth]{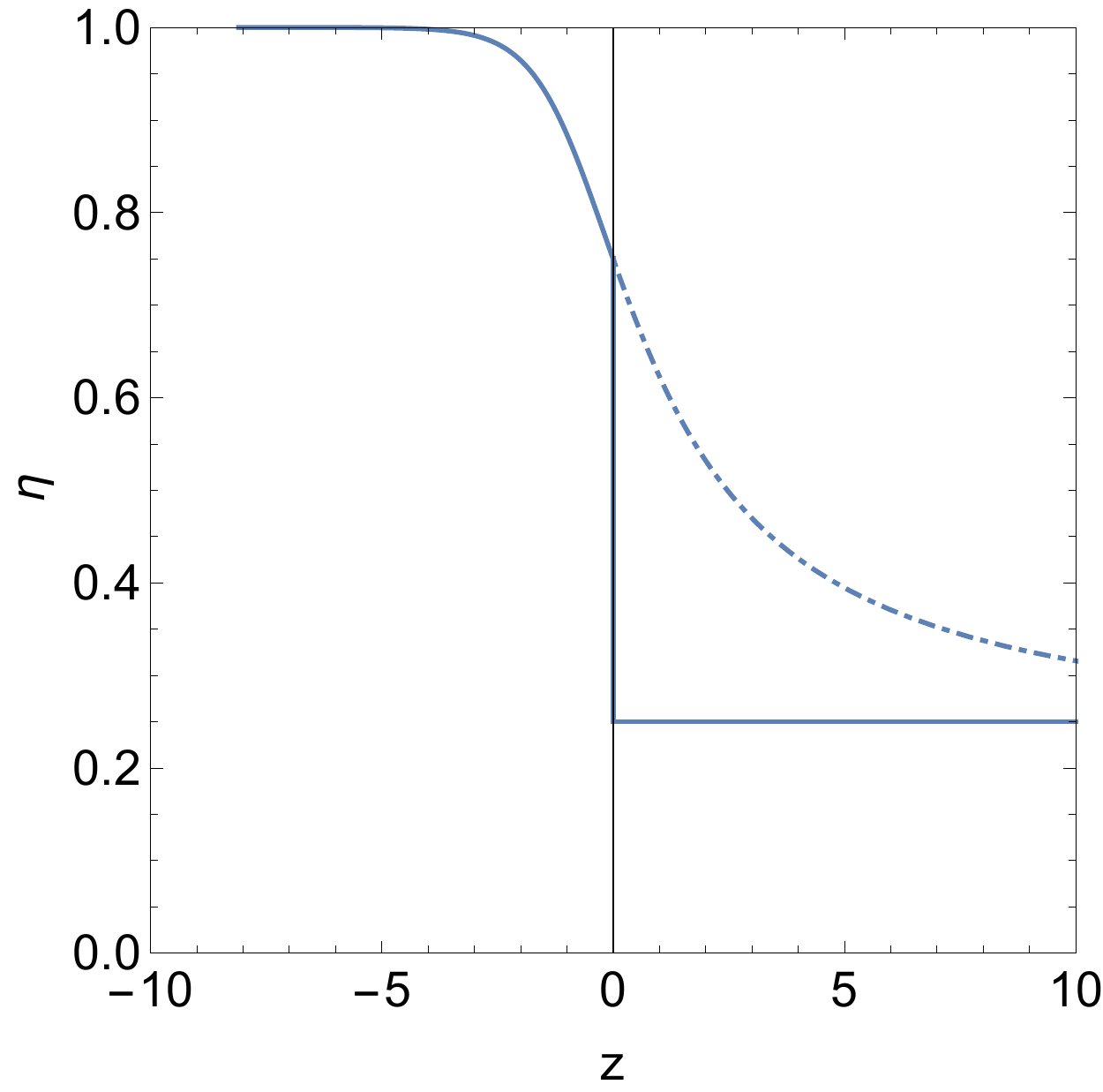}
\caption{  Evolution of the compression ration and the velocity in   CR-modified shock with $\gamma=5/3$. Iso-thermal jump is at $z=0$,   dot-dashed line corresponds to the absence of the isothermal jump, Eq. (\ref{sol}).
The iso-thermal jump connects states with $\eta_+ =3/4$ and $\eta_2 =1/4$ (compare, \eg, with \cite{1982A&A...111..317A} Fig. 5 and \cite{1981ApJ...248..344D}, Fig. 2). 
}
% \vskip -.5 truein
\label{zofeta}
\end{figure}

Dimensinalizing distance by  ${\kappa}/{ v_1^3}$, Eq. (\ref{main}) can be integrated
\ba &&
z= \log \left((1-\eta )^{-\frac{2 \eta _2}{\eta _2-1}} \eta ^{\frac{2}{\eta _2}+2}
   \left(\frac{1}{\eta _2-1}+2\right){}^{\frac{2}{\eta _2-\eta _2^2}} \left(1-\eta
   _0\right){}^{-\frac{2 \left(\eta _2+1\right)}{\eta _2}} \eta _2^{\frac{2 \eta _2}{\eta
   _0-1}} \left(\frac{\eta _2-\eta }{\eta _2-1}\right){}^{\frac{2}{\left(\eta _2-1\right)
   \eta _2}}\right)-\frac{2}{\eta }-\frac{2}{\eta _2-1}
   \nn  &&
   \rightarrow
   \log \left(\frac{1024\ 2^{2/3} (1-\eta )^{2/3} \eta ^{10}}{59049 \left(\eta
   -\frac{1}{4}\right)^{32/3}}\right)-\frac{2}{\eta }+\frac{8}{3}
   \label{sol}
   \ea
   where the integration constant as been chose so that the thermal jump is located at $z=0$, see Fig. \ref{zofeta}.

\section{Cosmic rays acceleration at the iso-thermal jump}

Let us first give relations for  the strong shock limit $M_1 \rightarrow \infty$. At the iso-thermal jump the sound speed is 
\be
c_s = \sqrt{\gamma T/m_p} = \frac{\sqrt{2} \sqrt{\gamma -1} \sqrt{\gamma }}{\gamma +1} v_1
\ee
At this point, on the upper branch the  parameters of the flow are
\ba && 
\eta_+ = 2/(\gamma+1)
\nn &&
v_+= \frac{2}{\gamma +1} v_1 = \frac{3}{4} v_1
\nn &&
M_+ = \frac{\sqrt{2}}{\sqrt{\gamma -1} \sqrt{\gamma }} = \frac{3}{\sqrt{5}}
\ea
While in the post-jump flow
\ba
 && 
\eta_2 = \frac{\gamma-1}{\gamma+1}
\nn &&
v_2= \frac{\gamma+1}{\gamma-1}v_1 = \frac{1}{4} v_1
\nn &&
M_2 = \frac{\sqrt{\gamma -1}}{ \sqrt{2 \gamma }} = \frac{1}{\sqrt{5}}
\ea
The compression ratio at the  isothermal jump is $ (\gamma-1)/2=1/3$, so that the expected spectrum of CRs is 
$p = 2\gamma/(3- \gamma) = 5/2$ \citep[\eg][]{1987PhR...154....1B}, steeper than conventional $2$ for $\gamma =5/3$.

\begin{figure}[h!]
 \vskip -.1 truein
 \centering
\includegraphics[width=0.99\textwidth]{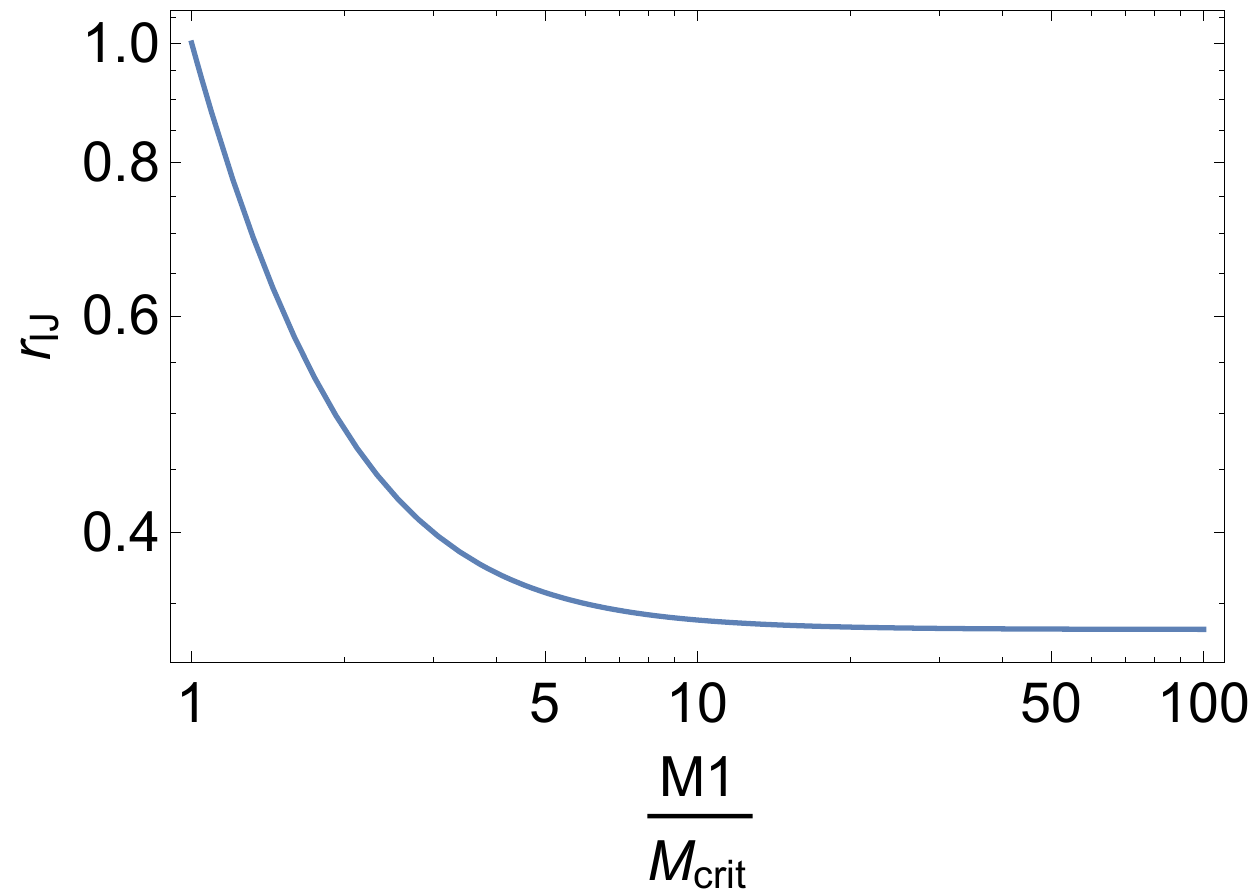}
\caption{ Inverse of the compression ration at the isothermal jump as a function of  the ratio of upstream Mach number to the critical one.}
% \vskip -.5 truein
\label{rij}
\end{figure}

For finite upstream Mach number the compression jump $r_{IJ}$ at the isothermal shock  is 
\be
r_{IJ}= \frac{ \eta_+} { \eta_-}= \frac{\left(3 \gamma ^2-4 \gamma +1\right) m_1^2-2 (\gamma -3) \gamma }{\gamma ^2-4 \gamma +(6 \gamma -2) m_1^2+3} \rightarrow 1/3,
\ee
where the last limit assumes $m_1 \gg1 $ and $\gamma=5/3$.
see Fig. \ref{rij}.

 \section{Discussion}
 In this work we point out that for sufficiently strong shocks, with Mach number $ M_1 >  \sqrt{\frac{3 \gamma -1}{(3-\gamma ) \gamma }}=3/\sqrt{5}=  1.34$,
 the energy diffusion induced by CRs modifies the global structure, creating a special kind of a discontinuity - an isothermal jump. At the isothermal jump temperature remains constant - hence, it is not a shock. On the other hand the flow does change from supersonic to subsonic, with a compression ratio of $3$. Thus, many models of diffusive shock acceleration that rely on the shock compression ratio are likely to remain valid for the isothermal jump as well.
 
We stress that the appearance of the  isothermal jump is, generally, independent on the particular form of CR feedback - it is the evolution of the flow towards the 
 isothermal jump that is affected by a particular feed-back mechanism. It is not that large conductivity makes the sub-shock isothermal -  even the minimal CR-diffusion leads to the formation of the iso-thermal jump (for super-critical shocks). 
 
 The density compression ratio at the iso-thermal jump of $3$ will lead to the spectral index of accelerated particles of $p=2.5$ \citep[in the limit of  test-particle experiencing   first-order Fermi process][]{1977DoSSR.234.1306K,1978ApJ...221L..29B}, somewhat steeper than the conventional values of $p=2$ for strong shocks with 
compression ratio of $4$. In fact, in  many settings the inferred spectra are closer to $p=2.5$ \citep[\eg, as discussed by][]{2012JCAP...07..038C}. At the nonlinear stage, when CRs start strongly affect the thermodynamic properties of the flow, the adiabatic index can decrease to $\gamma=4/3$, which would give compression ratio at the isothermal transition of $2/(\gamma -1) = 6$. 
 
 Our hydrodynamic approach naturally has severe limitations. Kinetic effects are likely to be important, especially for the very highest energy particles. But since the predicted spectrum is soft, the CR escape from the shock will not be important. 
 
 Finally, let us comment on the effects of \Bf.
Magnetic fields slightly modify the appearance of the isothermal jump. For strong perpendicular shocks
instead of (\ref{T}) we have
\be
T= (1-\eta) \eta m_p v_1^2 \left( 1 - \frac{1+\eta}{2 \eta^2}  \frac{1}{M_A^2} \right)
\ee
where $M_A = v_1 /v_A$ and $v_A$ is \Alfven velocity. Thus, correction is small for highly super-Alfvenic flows. Generally, for perpendicular shocks the effect of \Bf\ on the fluid flow can be completely absorbed into the definition of sound speed - which becomes the fast magnetosonic speed.

I would like to thank Damiano Caprioli, Donald Ellison and  Mikhail Malkov for discussions and organizers of the  workshop  ``Cosmic Accelerators'' at the Joint Space-Science Institute  where part of this work has been performed. This work had been supported by   NSF  grant AST-1306672, DoE grant DE-SC0016369 and NASA grant 80NSSC17K0757.

\bibliographystyle{apj}

  \bibliography{/Users/maxim/Home/Research/BibTex}

 \end{document}